# *Dynamic magnetic features of a mixed ferro-ferrimagnetic ternary alloy in the form of $AB_pC_{1-p}$*

Mehmet Batı[1,a]   Mehmet Ertaş[b]

[a]*Department of Physics, Recep Tayyip Erdoğan University, Rize, Turkey*

[b]*Department of Physics, Erciyes University, Kayseri, Turkey*

**Abstract**

Dynamic magnetic features of a mixed ferro-ferrimagnetic ternary alloy in the form of $AB_pC_{1-p}$, especially. The effect of Hamiltonian parameters on the dynamic magnetic features of the system are investigated. For this aim, an $AB_pC_{1-p}$ ternary alloy system was simulated within the mean-field approximation based on a Glauber type stochastic dynamic and for simplicity, *A*, *B* and *C* ions as $S_A = 1/2$, $S_B = 1$ and $S_C = 3/2$, were chosen respectively. It was found that in our dynamic system the critical temperature was always dependent on the concentration ratio of the ternary alloy.

### 1- Introduction

Mixed-spin Ising systems provide a good model for studying a ferro-ferrimagnetic ternary alloy. Ternary alloy in the form of $AB_pC_{1-p}$ composed of Prussian blue analogs [1] have been the topic of much research because of their interesting magnetic behaviors, such as photo induced magnetization, charge-transfer-induced spin transitions, the existence of compensation temperatures and hydrogen storage capacity [2-11]. It has been widely studied in literature by various methods in equilibrium statistical physics, such as exact recursion relations (ERR) on the Bethe lattice, mean-field approximation (MFA), Monte Carlo (MC) simulations and effective-field theory (EFT) with correlations [12-29]. In the studies, ternary alloy in the form of $AB_pC_{1-p}$ has been modelled with different magnitudes of spins, namely *A*, *B* and *C*. For instance, the magnetic features of a mixed ferro-ferrimagnetic ternary alloy in the form of $AB_pC_{1-p}$ consist of three different metal ions with ternary Ising spins (1/2, 1, 3/2) [12-16]; (3/2, 1, 1/2) [17]; (1/2, 1, 5/2) [18]; (1/2, 3/2, 5/2) [19], (1, 3/2, 5/2) [20-23], (3/2, 1, 5/2) [24-26] and with Ising spins (3/2, 2, 5/2) [27-29]. It is worth noticing that there are many experimental studies on Prussian blue analogs [30–36].

mehmet.bati@erdogan.edu.tr

Generally, dynamic property investigations of $AB_pC_{1-p}$ ternary alloys are difficult because of their structural complexity. Despite, the equilibrium magnetic properties of a mixed ferro-ferrimagnetic $AB_pC_{1-p}$ ternary alloy having been studied in detail, there are only two studies within our best knowledge about the dynamic properties of the $AB_pC_{1-p}$ ternary alloy system containing spin (3/2, 1, 5/2) [37, 38]. The aim of this study was focused on studying the dynamic magnetic properties of a ternary alloy system. Dynamic phase transitions (DPT) were obtained and the effect of Hamiltonian parameters on the dynamic properties of the system were investigated. For this aim, an $AB_pC_{1-p}$ ternary alloy system was simulated by utilizing the MFA based on Glauber type stochastic dynamic and for simplicity, A, B and C ions as $S^A = 1/2$ $S^B = 1$ and $S^C = 3/2$ were chosen, respectively. It is worth noting that the DPT has been extensively studied theoretically [39-45] and experimentally [46-50] for different systems during the last decades.

The paper is organized in the following way: Section II describes the models and its formulations. Section III presents the numerical results. Finally, the conclusion is given in Sec. IV.

## 2- Model formulation

A mixed ferro-ferrimagnetic $AB_pC_{1-p}$ ternary alloy system was considered on two interpenetrating square sublattices. One of the sublattices only has spins $S^A = \pm 1/2$ as the other sublattices have spins $S^B = \pm 1, 0$ or $S^C = \pm 3/2, \pm 1/2$. A sketch of the present model can be seen in Fig. 1. The Hamiltonian can be written as follows:

$$\mathcal{H} = -\sum_{<ij>} S_i^A \left[ J_{AB} S_j^B \xi_j + J_{AC} S_j^C (1-\xi_j) \right] - h(t) \left[ \sum_i S_i^A + \sum_j \left( S_j^B \xi_j + S_j^C (1-\xi_j) \right) \right], \qquad (1)$$

where $<ij>$ shows a summation over all pairs of the nearest-neighboring sites of different sublattices and $J_{AB} > 0$ and $J_{AC} < 0$ (model the ferro-ferrimagnetic interactions) are the nearest-neighbor exchange constants. $h(t)$ is the oscillating external magnetic field and is described by $h(t) = h_0 \cos(wt)$, where $h_0$, $w$ and $t$ are the time, amplitude and angular frequency. $\xi_j$ is distributed random variables and it takes the value of unity or zero, according to whether site $j$ is filled by an ion of B or C, respectively. So, $\xi_j$ is described by

$$P(\xi_j) = p\delta(\xi_j - 1) + (1-p)\delta(\xi_j), \qquad (2)$$

where $p$ and $(1-p)$ are the concentration of B and C ions, respectively. A mixed ferro-ferrimagnetic $AB_pC_{1-p}$ ternary alloy system is in contact with an isothermal heat bath at an absolute temperature $T_{abs}$ and evolves according to the Glauber-type stochastic process at a rate of $1/\tau$. From the master equation associated to the stochastic process, it follows that the average magnetization satisfies the following equation [39-45],

$$\tau \frac{d}{dt}\langle S^A \rangle = -\langle S^A \rangle + \frac{1}{2}\tanh\left[\beta\left(J_{AB}p\sum_j S_j^B + J_{AC}\sum_j S_j^C(1-p) + h_0\cos(wt)\right)\right], \quad (3a)$$

$$\tau \frac{d}{dt}\langle S^B \rangle = -\langle S^B \rangle + \frac{3\sinh\left[\beta\left(J_{AB}p\sum_i S_i^A + ph_0\cos(wt)\right)\right]}{2\cosh\left[\beta\left(J_{AB}p\sum_i S_i^A + ph_0\cos(wt)\right)\right]+1}, \quad (3b)$$

$$\tau \frac{d}{dt}\langle S^C \rangle = -\langle S^C \rangle + \frac{3\sinh[1.5(1-p)\beta x] + \sinh[0.5(1-p)\beta x]}{2\cosh[1.5(1-p)\beta x] + 2\cosh[0.5(1-p)\beta x]}, \quad (3c)$$

where $x = J_{AC}\sum_i S_i^A + h_0\cos(wt)$. Using the mean-field theory; the dynamic mean-field approximation equations are obtained as follows

$$\Omega\frac{d}{d\zeta}m_A = -m_A + \frac{1}{2}\tanh\left[\beta\left(J_{AB}z_{AB}m_A p + J_{AC}z_{AC}m_C(1-p) + h_0\cos(\zeta)\right)\right], \quad (4a)$$

$$\Omega\frac{d}{d\zeta}m_B = -m_B + \frac{3\sinh\left[\beta\left(J_{AB}z_{BA}m_A p + ph_0\cos(\zeta)\right)\right]}{2\cosh\left[\beta\left(J_{AB}z_{BA}m_A p + ph_0\cos(\zeta)\right)\right]+1}, \quad (4b)$$

$$\Omega\frac{d}{d\zeta}m_C = -m_C + \frac{3\sinh[1.5(1-p)\beta y] + \sinh[0.5(1-p)\beta y]}{2\cosh[1.5(1-p)\beta y] + 2\cosh[0.5(1-p)\beta y]}, \quad (4c)$$

where $y = J_{AC}z_{CA}m_A + h_0\cos(\xi)$, $\xi = wt$, $\Omega = \tau w$ and $\Omega = 2\pi$, $z_{AB}$, $z_{BA}$, $z_{AC}$ and $z_{CA}$ are taken 4 for a square lattice.

$$M_i = \frac{1}{\tau}\int m_i(\xi)d\xi \quad (5)$$

where $i = A, B$ and $C$. In other words, $M_A$, $M_B$ and $M_C$ correspond to the dynamic order parameters of the magnetic components $A$, $B$ and $C$. The total magnetization of the system is

$$M_T = \frac{(M_A + M_B + M_C)}{2} \tag{6}$$

The physical parameters have been scaled in terms of $J_{AB}$. For example, reduced temperature and field amplitude are respectively defined as $T = \frac{k_B T_{abs}}{J_{AB}}$, and $h = \frac{h_0}{J_{AB}}$, throughout the

### 3- Results and Discussion

The effects of the concentration ratio $p$ and the exchange interaction ratio $R$ ($\frac{|J_{AC}|}{J_{AB}}$) on dynamic magnetization and DPT of the ternary alloy have been examined. It should be noted that the $p = 0$ case corresponds to a ferrimagnetic mixed spin-1/2 and spin-3/2 system while for $p = 1$, corresponds to a mixed spin-1/2 and spin-1 ferromagnetic system. The phase diagram of the ternary alloy in $(R - T_C)$ and $(p - T_C)$ planes are shown in Fig. 2 and Fig. 3, respectively. In these figures, upper graphs are plotted for $h = 0.1$ and the lower ones are plotted for $h = 0.5$. The critical temperature value (phase transition temperature) is a little decreased with increasing $h$. ($T_C = 2.71$ for $h$=0.1, $T_C = 2.67$ for $h$=0.5) The reason for this situation is that, the higher field amplitude becomes dominant against the ferromagnetic and antiferromagnetic nearest-neighbor bonds. It can be seen that from the figures that $T_C$ increases as $R$ increases and the $T_C$ values do not change with $R$ for $p = 1.0$. Because the system become an $AB$ alloy, there is no $AC$ interaction. Therefore the system becomes independent from $R$.

In this section, the effects of $p$ and $R$ on the magnetization of a ternary alloy of the type $AB_pC_{1-p}$ are discussed. Fig. 4 shows the total magnetization chancing with scaled temperature for $R$=0.5, 1.0 and $R$=2.0 values. It is again seen from the figures that all the total magnetization curves merge at a unique transition temperature for $p = 1.0$. For the $p$ =0.0 case, $Tc$=0 at $R$ =0.0 (see Fig. 1 and 2) and the dynamic critical temperature of the system increase with an increasing of $R$. For $p$=0.25, the antiferromagnetic exchange interaction between the $A$ and $C$ magnetic components becomes effective in the system for larger R values. In other words, the $J_{AC}$ interaction becomes dominant and because $J_{AC}$ is negative, the $M_T$ results are negative. It is noted that the saturation values of magnetization increase for $p$=0.25 and decrease for $p$=0.75 with the increasing of $R$. A second order phase transition occurs in the system for $R$=0.5 and $R$=1.0. But for $R$=2, first the system gives the first order phase transition and then the second order phase transition occurs. $M_T$ decreases as $R$ increases in the range $0.5 < R < 1$ at $p = 0.5$ and after $R > 1$ (*AC* interaction is more dominant), as $R$ increases, $M_T$ orientation increases by changing.

In Fig. 5 the results have been depicted for $R=1.0$. As expected, the sign of the $M_A$ magnetization is negative ($M_A = -1/2$) $M_B = 0.0$ and $M_C = 3/2$ at $T=0$ for $p=0.0$. For $0.0<p<1.0$ the sign of the $M_C$ magnetization is negative ($M_C = -3/2$), $M_A = 0.5$ and $M_B = 1.0$ at $T=0$. Lastly, for $p=1.0$ $M_A = 1/2$, $M_B = 1.0$ and $M_C = 0$ at $T=0$. There are no dynamic magnetic properties investigation of Ternary alloy for spin (½, 1, 3/2). However, there is an investigation for spin (3/2,1,5/2) [37]. Very interesting behaviors with the variations of scaled temperature, concentrations of the spins and ratio of the bilinear interactions are found in agreement with the literature. An interesting feature that has been observed in equilibrium systems is the existence of a special interaction ratio $R_C=|J_{AC}|/J_{AB}$ where the critical temperature of the system becomes independent of the concentration ratio $p$ of the ternary alloy [25]. However, in the dynamic system it has been observed that critical temperature is always dependent on the concentration ratio $p$ of the ternary alloy. Theoretically, this behavior was first reported firstly in literature by Vatansever [37]. Dynamic behaviors of ternary alloys have not yet been studied experimentally. Actually, there are only a few experimental studies about dynamic phase transition in literature [51-53], such as, time dependent magnetic behavior of uniaxial cobalt films in the vicinity of the dynamic phase transition (DPT) as a function of an oscillating magnetic field experimentally studied by Berger and coworkers [51].

4- **Conclusions**

In this paper, the dynamic magnetic features of the mixed ferro-ferrimagnetic ternary alloy of the type $AB_pC_{1-p}$ on a square lattice, composed of spins 1/2, 1, and 3/2 has been studied. A ternary alloy system was simulated within the mean-field approximation based on Glauber type stochastic dynamics. The concentration ratio was obtained as well as the exchange interaction ratio dependence of critical temperature. It was found that the critical temperature could be controlled by the concentration ratio and the exchange interaction ratio. In particular, it has been found that the critical temperature is always dependent on the concentration ratio $p$ of the ternary alloy in the dynamic system. It is believed that this study will shed light on future experimental and theoretical investigations of mixed ferro-ferrimagnetic compounds in the presence of a time dependent external magnetic field.

**List of the Figure Captions**

Fig. 1: (color online) Schematic representation of ternary alloy spin model the type of $AB_pC_{1-p}$ defined on a square lattice. *A*, *B* and *C* represented in yellow, blue and red respectively. Interaction between A and B sites is ferromagnetic (FM) and *A* and *C* antiferromagnetic (AFM).

Fig. 2: Dynamic phase diagrams of the system in a $(R - T_C)$ plane for the different values of $p$ with $h = 0.1$ and $h = 0.5$.

Fig. 3: Dynamic phase diagrams of the system in a $(p - T_C)$ plane for the different values of $R$ with $h = 0.1$ and $h = 0.5$.

Fig. 4: (color online) Scaled temperature dependence of total magnetization for different values of $p$ with $R = 0.5$ (upper figure), $R = 1.0$ (middle figure) and $R = 2.0$ (lower figure).

Fig. 5: (color online) Scaled temperature dependence of magnetization ($M_A$, $M_B$ and $M_C$) for different values of $p$ with $R = 1.0$.

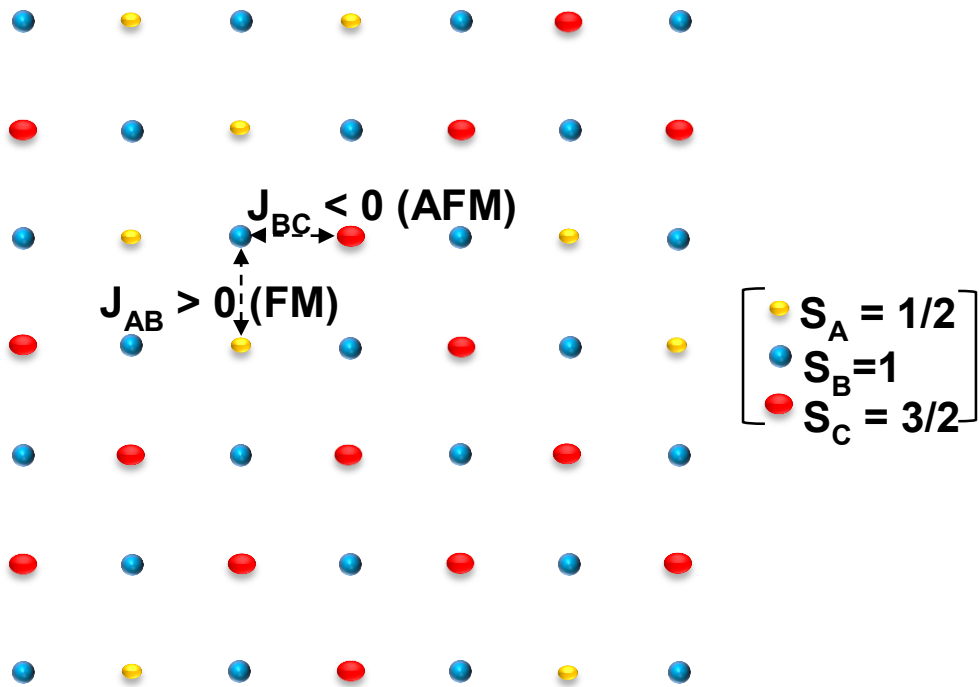

Fig. 1:

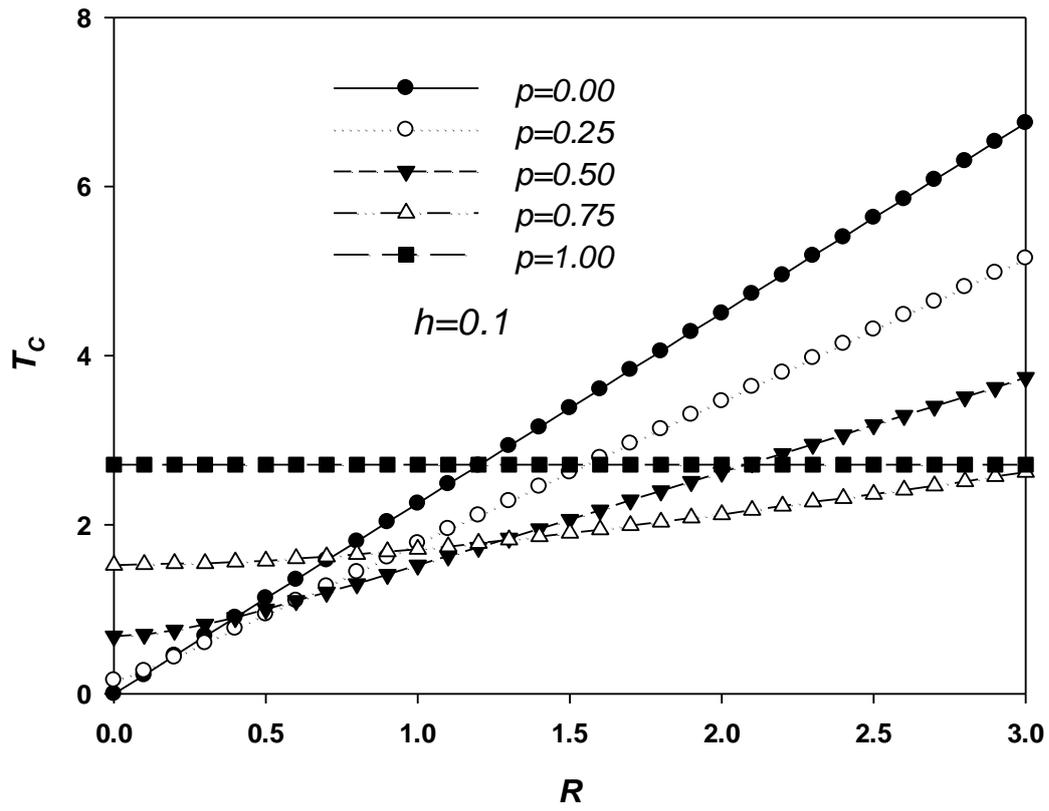

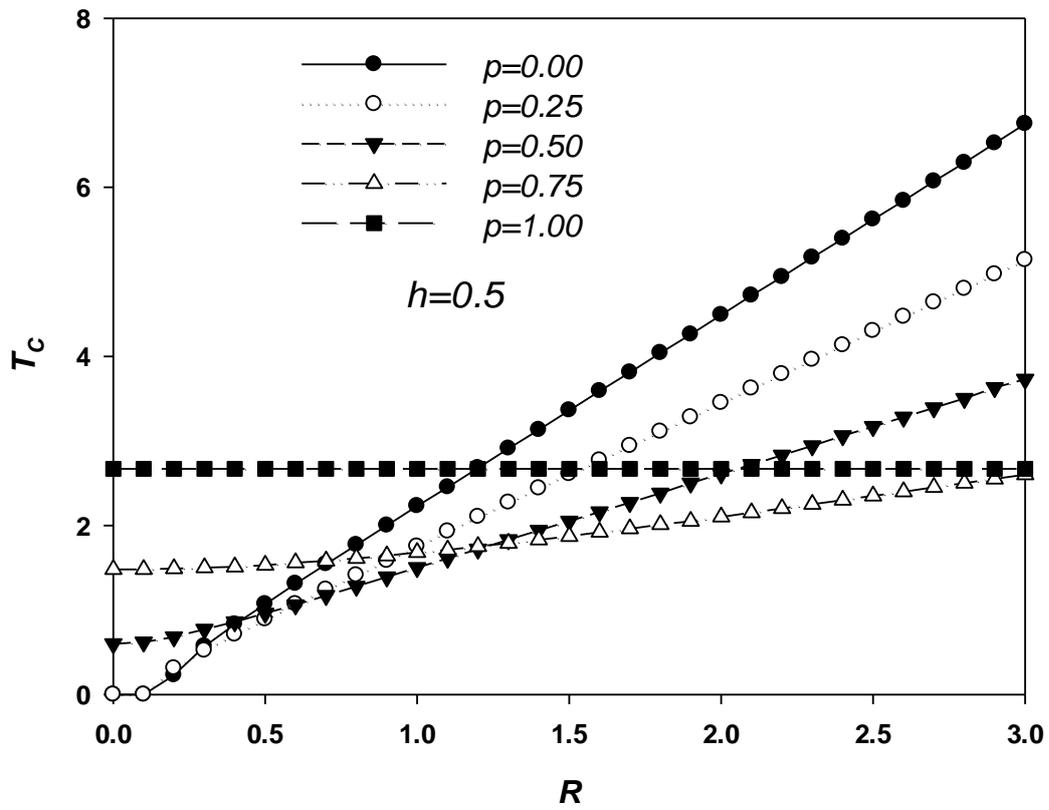

Fig. 2:

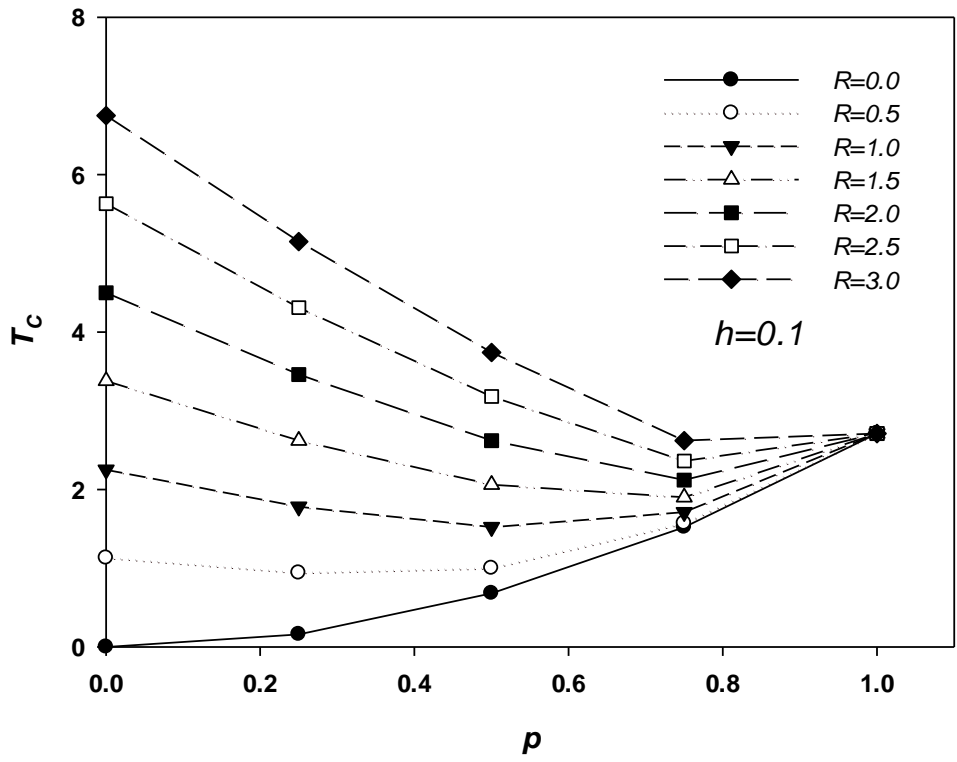

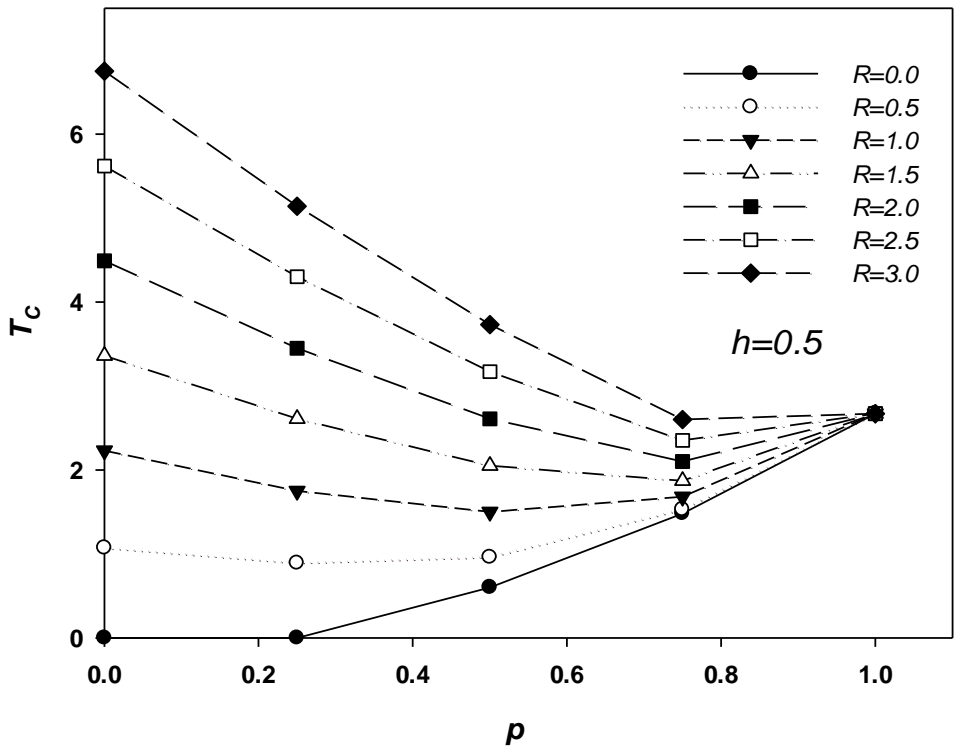

Fig. 3:

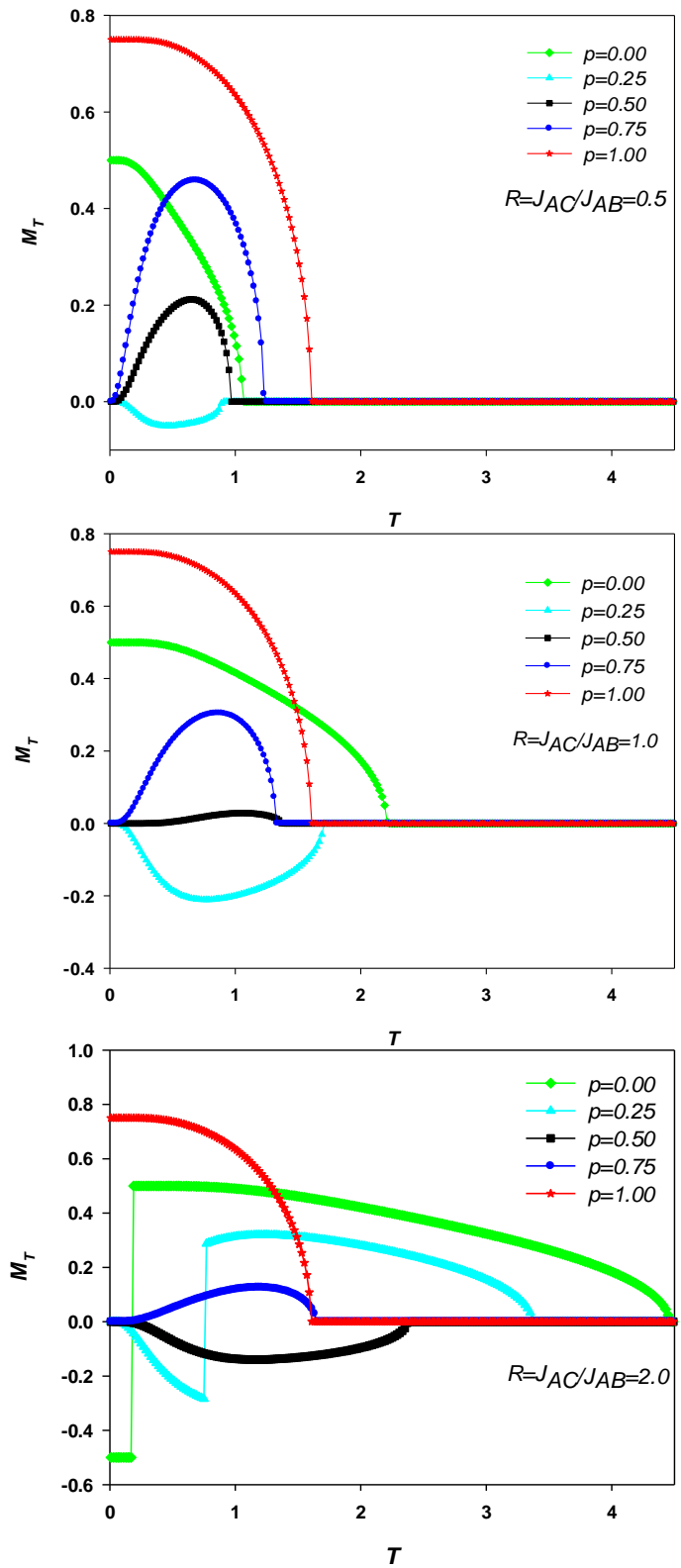

Fig. 4:

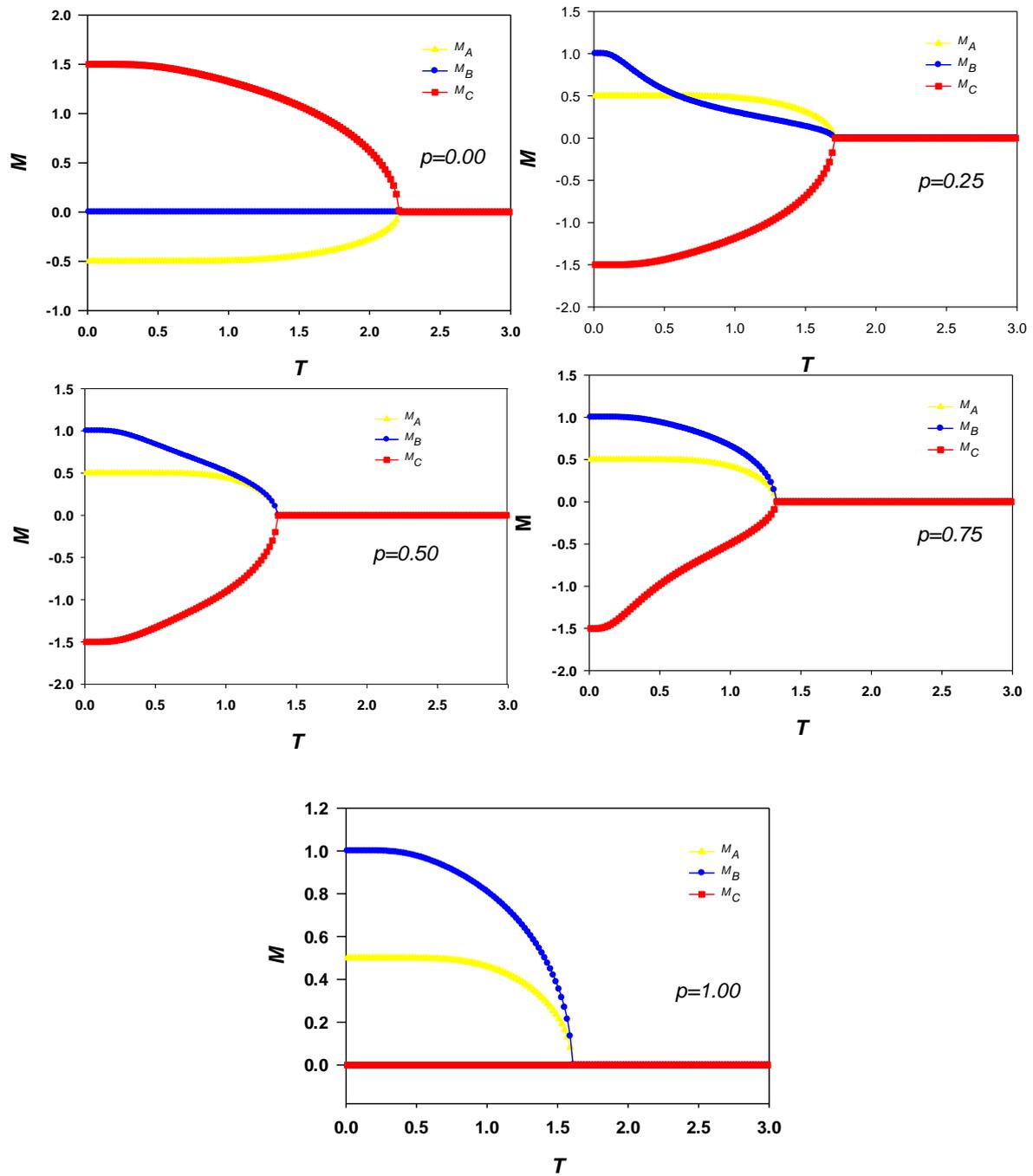

Fig. 5: